# SEARCHING FOR THE PROPERTIES OF NUCLEAR MATTER USING PROTON-CARBON AND DEUTERON-CARBON COLLISIONS AT 4.2 A GeV/c

M. AJAZ*,†, M. K. SULEYMANOV*,‡, K. H. KHAN*,§ and A. ZAMAN*,¶

*COMSATS Institute of Information Technology,
Park Road, Islamabad, 44000, Pakistan
†Department of Physics, Abdul Wali Khan University,
Mardan, 23200, Khyber Pukhtoonkhwa, Pakistan
‡Joint Institute for Nuclear Research, Dubna, Moscow region, Russia
†Muhammad.Ajaz@cern.ch
‡mais_suleymanov@comsats.edu.pk
§rph_f06_023@comsats.edu.pk
¶ali_zaman@comsats.edu.pk

## Abstract

The present work reports the use of nuclear transparency effect of protons in proton and deuteron carbon interactions at 4.2 A GeV/c to get information about the states of nuclear matter. The "half angle" technique is used to extract the information on nuclear transparency. The results are compared with Dubna version of Cascade model. The average values of multiplicity, momentum and transverse momentum of protons are analyzed as a function of the number of identified protons in an event. We observed some evidence and trends in the data which could be considered as transparency effect. Analysis of the results shows that the leading effect is the basis of the observed transparency. Some contribution to the observed effect could be the existing short range correlations and the scaling power law $s^{-N}$, for exclusive two–body hard scattering.

## I. INTRODUCTION

Nuclear Transparency (NT) effect is an important phenomenon which could reflect some particular properties of a medium when connected to the dynamics of hadron-nuclear (hA) and nuclear–nuclear (AA) interactions. Transparency in experiments has already been defined by a number of researchers [1-7]

The study of the nuclear transparency (NT) effect in hA and AA collisions was done by P. L. Jain et al., [6] using "half angle" ($\theta_{\frac{1}{2}}$) technique. The angle $\theta_{\frac{1}{2}}$ is defined as an angle which divides the particle multiplicity into two equal parts in a nucleon-nucleon (NN) interaction. They studied the behavior of s-particles (the particles with $\beta>0.7$ in the emulsion experiments) as a function of g-particles (the particles with $0.23 \leq \beta \leq 0.7$). A. I. Anoshin, et. al, [1] studied the average multiplicity of pions as a function of the number of identified protons ($N_p$) and found that the average multiplicity of pions is independent of $N_p$. The authors observed that the average momentum of these particles is a decreasing function of $N_p$ and concluded that the observed behavior could not be considered as total transparency. In our case we used the half angle technique and defined the NT as an effect at which the characteristics of hadron-nucleus and nucleus-nucleus collisions do not depend on the number of identified protons ($N_p$), because $N_p$ is connected with baryon density of matter. The idea of nuclear transparency, for the first time, at low energy was given by Bethe [8] in 1940. At high energies NT is studied intensively at BNL [2, 9, 10], SLAC [11, 12] and JLab [13] with protons as a probe for studying the effect. Later mesons (qq´) were used to see an increase in the NT [3, 14-20]. That gives substantial and clear signature for the increase in the NT in most of the cases. The critical change in the NT is considered as an important effect to get the information on particular properties of strongly interacting matter as well as the Quark Gluon Plasma (QGP) [21, 22].

## II. THE METHOD

Using pp collision we determined the values of the $\theta_{½}$ to be $\theta_{½} = 25^o$. The $\theta_{½}$ divides the particles into the incone and outcone particles. So the particles with $\theta < \theta_{½}$ are the incone particles and those with $\theta > \theta_{½}$ are the outcone particles. We defined the NT as an effect at which the characteristics of protons in $p^{12}C$ and $d^{12}C$ do not depend on $N_p$. $N_p$ is used to fix the centrality of collisions because $N_p$ connects with baryon density of matter. Finally the results are compared with the data coming from Dubna version of cascade model [23-27].

## III. THE CASCADE MODEL

Cascade model is the most popular model which is used to describe the general features of relativistic nucleus-nucleus collisions. It is an approach based on simulation using Monte-Carlo techniques and is applied to situation where multiple scattering is important.

The basic assumptions and procedures of the cascade model are given by K. K.Gudima et al. [23] and A. Boudard et. al [24]. A brief description of the model is presented here. The cascade model does not include any medium or collective properties and that each colliding nuclei is treated as a gas of nucleon bound in a potential well. The model is used for multiple scattering based on Monte-Carlo simulation techniques. Particle production takes place when a projectile interacts with target after absorbing momentum. The produced particles interact elastically or inelastically with other nucleons and produce new particles. This process continues till the moving particle either leaves the nucleus or is absorbed. A diffused distribution of nuclear density and nuclear potential is assumed and correlation between nucleons in the ground state taken into account. In Monte Carlo study of the time evolution of two interacting nuclei, the inter dependence of target and cascade are achieved through correlations of density of the nucleons from colliding nuclei in intra nuclear collisions. The Pauli principle and the energy momentum conservation are obeyed in each inter-nuclear interaction. The remaining excited nuclei, after the cascade stage are described by the statistical theory in the evaporation approximation. In comparison with the experimental data we used the same conditions to both sets of events.

## IV. THE EXPERIMENT

We used the experimental data obtained from the 2-m propane bubble chamber of the laboratory of high energy of the Joint Institute for Nuclear Research (JINR), Dubna, Russia. The data was obtained on the basis of processing stereo photographs. The chamber was exposed to beams of protons and deuterons accelerated to a momentum of 4.2 A GeV/c which was kept in a 1.5 Tesla magnetic field at the Dubna Synchrophasotron. The detail discussion on the interaction mechanism is given by H. N. Agakishiyev et al. [28]. Practically all secondary identified particles emitted at a $4\pi$ solid angle were detected in the chamber. All negative particles, except identified electrons, were considered as $\pi^-$ - mesons. The contaminations caused by misidentified electrons and negative strange particles do not exceed 5% and 1%, respectively. The threshold momentum for pion registration used was 70 MeV/c below which the pion could not be identified because of their short range in the chamber. The protons were selected by the statistical method applied to all positive particles with momentum of 150 MeV/c because of their short range in the chamber (slow protons were identified with p ≤ 700 Mev/c by ionization in the chamber). The average error in measuring angles of the secondary particles was $0.8^o$, while the mean relative error in determining momenta of the particles from the curvature of a track in the magnetic field was 11%. The corrections to account for losses of particles emitted under large angles to the object plane of the camera were introduced: They amounted to ~3% for protons with momenta $p_{lab}$ > 300 MeV/$c$ and ~15% for slow protons with $p_{lab}$ < 300 MeV/$c$ [29]. In this experiment, we used 12757 pC , 9016 dC , interactions at a momentum of 4.2 A GeV/c ( for methodical details see [30]). In the case of cascade code we used 50000 pC- and dC-interactions under the same conditions.

## V. RESULTS

The values of average multiplicity $<n>_{pC}$, average momentum $<p>_{pC}$, and average transverse momentum $<p_T>_{pC}$ of protons from experimental data and Cascade model in pC collisions at 4.2 A GeV/c as a function of $N_p$ are shown in Fig. 1. The value of $\theta_{½}$ used is $25^o$. The experimental results are shown by geometrical symbols (■) for $<n>_{pC}$, (□) for $<p>_{pC}$, (▲) for $<p_T>_{pC}$. The results from the cascade model are shown by solid line for $<n>_{pC}$, broken line for $<p>_{pC}$ and dash line for $<p_T>_{pC}$.

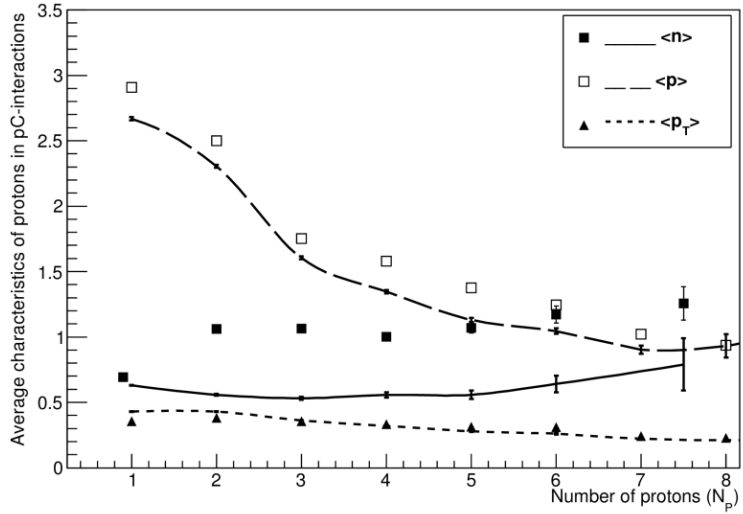

FIG. 1. Behavior of the average characteristics (average multiplicity, average momentum, average transverse momentum) of protons in $p^{12}C$- interaction at 4.2 A GeV/c as a function of number of identified protons ($N_p$) at half angle ($\theta_{½}$) = $25^0$. Geometrical symbols show the results of the experimental data whereas lines show the results of the cascade model as are given in the legend against each characteristic.

The behavior of $<n>_{pC}$ doesn't depend on $N_p$ in the region of $N_p$ = 2-8 with a very slight positive slope (+0.04±0.02 for experimental data and +0.06±0.01 for cascade model). The results in the two cases are fitted using linear function $<n>_{pC} = A + B*N_p$. Where A and B are free parameters with B the slope of the line. The $N_p$-dependence demonstrates clearly transparency for the protons with $<n>_{pC} \approx 1$. Thus we can say that the experimental data demonstrate clearly transparency for the protons with $\theta_{½}$= $25^0$. The experimental behavior of $<n>_{pC}$ is suppressed by the Cascade Model (see Fig. 1). This can be explained with some mixture of the fast $\pi^+$-mesons among these protons in the experiment. The fast $\pi^+$-mesons could appear as a result of proton exchange reactions when proton in the wounded nucleon transform into a neutron and fast positive pions are emitted through the reactions p→n + $\pi^+$ [31]. According to the law of conservation of momentum the $\pi^+$-meson gets most of the energy as their kinetic energy.

The behavior of $<p>_{pC}$ as a function of the $N_p$ is represented by □, whereas the corresponding result from cascade model is shown by broken line. There are two regions for the behaviors of the $<p>_{pC}$. In the first region ($N_p$ =1-3) the values of $<p>_{pC}$ decreases sharply and in the second one the values of $<p>_{pC}$ decreases slowly with $N_p$. No transparency is observed in this case and the code data gives almost the same behavior of sharp decrease in the first region and slow decrease in the second region for experimental results. The values of the cascade results are again less than that of the experimental one because of the reason explained above in terms of the increase in the number of protons due to the mixture of fast $\pi^+$-mesons in a charge exchange reaction.

The values for the $<p_T>_{pC}$ as a function of the $N_p$ for the experimental and code data are given in Fig. 1 by ▲ and dash line respectively. The cascade model supports the results of the experimental data very well. A linear function ($<p_T>_{pC} = A + B*N_p$) is used to fit the two results which gives the value of slope to be negative (-0.026±0.004). It means that the value of $<p_T>_{pC}$ is also a decreasing function of $N_p$.

Our first claim of the observed transparency for the incone protons' average multiplicity could be explained in terms of leading effect [32]. Leading particles are projectiles which could give-up a part of their energy during interaction. The particles will have maximum energy in an event and would be identified in an experiment as incone particles due to their high energy /low angle. Having high energy they are able to pass through the medium without losing a large fraction of their initial energy, making the medium transparent to them. This explanation is based on the fact that although the values of average multiplicity of the particles remain the same but their average momentum and average transverse momentum have been decreased. "Other reasons that could contribute to the observed transparency would be the existing of the short range correlations (SRC) [33], and the scaling power law $s^{-N}$ (where N+2 is the total number of elementary constituents in the initial and final state), for exclusive two–body hard scattering predicted by Matveev et al. [34], Brodsky and Farrar [35, 36] and Brodsky and Hiller [37]. Earlier analyses [38-40] assumed that the SRCs would be isospin-independent, with equal probability for pp, np and nn pairs to have hard interactions and generate high-momentum nucleons. But, the measurements of two-nucleon knockout showed that these correlations are dominated by np pairs [41, 42] due to the fact that the bulk of the high-momentum nucleons are generated via the tensor part of the N–N interaction rather than the short-range repulsive core [43, 44]. Experimentally, the constituent counting rule proves to be fairly successful, from the scattering of hadrons on protons to photoproduction of mesons [45, 46] to reactions involving light nuclei, like the photodisintegration of deuterons studied at Jefferson Lab [47, 48] . The computation of the experiments N of the scaling power in the AFS BNL experiments E838 (Ep = 5.9 GeV/c) and E755 (Ep = 9.9 GeV/c) are reported by [49]. The constituent counting predicts N = 10 for the pp→pp. Due to SRC the distance between inner nuclear nucleons decreases and the projectile could interact with several nucleons simultaneously (like with coherent nucleon tube [50]), which could give some additional transparency too. The density of the nucleus has been important in explaining the presence of high momentum nucleons arising from the SRCs. SRC is an important ingredient of the dynamics of nuclei which increases[51] the nuclear transparency- an important phenomena, connected with dynamics of hadron-nuclear and nuclear–nuclear interactions which could be used to get some information about particular properties of the medium.. We want to investigate the SCRs contribution in transparency with two particles correlation function ($R_{12}$ or $C_{12}$). This will be a complex investigation because with increase in the number of identified protons, the statistics of the data decrease reaching minimum at high values of $N_p$, where the transparency is more important for studying the properties of the nuclear matter."

Fig. 2 shows the average characteristics (average multiplicity ($<n>_{dC}$), average momentum ($<p>_{dC}$), and average transverse momentum ($<p_T>_{dC}$)) of incone protons from the experimental data and cascade model in deuteron-carbon (dC) -interactions at 4.2 A GeV/c as a function of $N_p$. Geometrical symbols are used for experimental data representation whereas lines are used for the results of cascade model. The behavior of $<n>_{dC}$ at $\theta_{½}=25^0$ is having a slight positive slope (0.09±0.02) as was the case in pC-interactions. The slope of the line is obtained from the result of fitting parameter B by linear function $<n>_{dC} = A+B*N_p$. The $N_p$-dependence demonstrates some transparency for these protons in the region of $N_p \geq 3$ with values greater than 1 as in comparison with Fig. 1 where the values were equal to 1. The result of cascade model shows that the values of the $<n>_{dC}$ are less than 1 as observed in the previous case of Fig. 1. Here again the model underestimates the results of experimental data. In this case the value of the $<n>_{dC}$ is large compared to that of the $<n>_{pC}$ because of the following charge exchange reaction n→p + $\pi^-$ [31]. In the two leader nucleon (neutron and proton in the projectile deuteron) the neutron can convert to a proton emitting a $\pi^-$-meson. The newly produced proton increases the average multiplicity as well as the average momentum of these leader protons in case of dC- interaction.

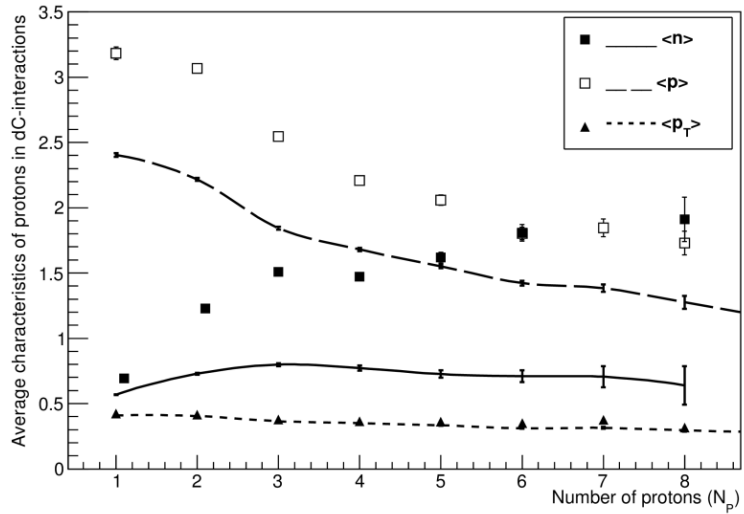

FIG. 2. Behavior of the average characteristics (average multiplicity, average momentum, average transverse momentum) of protons in $d^{12}C$-interation at 4.2 A GeV/c as a function of number of identified protons ($N_p$) at half angle ($\theta_{½}$) = $25^0$. Geometrical symbols show the results of the experimental data whereas lines show the results of the cascade model as are given in the legend against each characteristic.

The values of $<p>_{dC}$ as a function of the $N_p$ for the experimental and code data are given in Fig. 2 represented by open square (□) and broken line respectively. The values of $<p>_{dC}$ decreases in both cases with $N_p$. The cascade model fails to explain the results of the experimental data in this case because of the reason explained above. The neutron in the projectile deuteron can convert to a fast leading proton which is a reason of the boost in the average multiplicity and average momentum of these protons in the case of the dC- interactions.

Finally the values of $<p_T>_{dC}$ as a function of $N_p$ for experimental data and cascade model are shown by the same representation as before. Cascade model reproduce the results very well and is a decreasing function of $N_p$ in the two cases. The reason for the slight decrease in the case of $<p_T>$ is easy to understand as this variable is slowly varying function of many kinematical parameters.

It is important to mention that increasing mass of the projectile shifts the transparency towards high value of $N_p$. Beside this the results of the pC- interaction are well reproduced by the dC-interaction. The difference in the two results is due to the charge exchange reaction as explained above.

The same results were obtained using half angle technique in $\pi^{-12}C$-interaction at 40 GeV [52]. It was observed that the average multiplicity of $\pi^-$-mesons is independent of the $N_p$. It was also observed that the momentum of theses pion decreased with $N_p$ and declared that the behavior is not total transparency. The transparency observed [52] may also be due to the leading effect, because in this particular case $\pi^-$-mesons are the projectile particles. We produce similar results with protons and deuteron as projectile instead of pions and explained the results in terms of the leading effect transparency.

## VI. DISCUSSION AND CONCLUSION

Looking for the properties of the nuclear matter we studied the average characteristics of protons using $p^{12}C$- and $d^{12}C$- interactions at 4.2 A GeV/c. Nuclear transparency effect is considered as a signal of the appearance of different states of the strongly interacting matter. In this experimental investigation we observed some signal on the transparency of these protons. We used the experimental data of the JINR, Dubna Synchrophasotron having the $4\pi$ geometry measurement. We were able for the first time to use simultaneously 5 parameters: half angle; number of identified protons; the average multiplicity, average momentum and average transverse momentum for protons. We observed that the average multiplicity of protons in pC and dC interaction shows some evidence of transparency. Since the average momentum and average transverse momentum of these protons is a decreasing function of $N_p$ therefore it is concluded that the medium is not completely transparent. Analysis of the results shows that a reason of the transparency could be the leading effect. We believe that some contribution to the observed effect could give the existing short range correlations (SRC) and the scaling power law $s^{-N}$, (where N+2 is the total number of elementary constituents in the initial and final state), for exclusive two–body hard scattering. Due to SRC the distance between inner nuclear nucleons decreases and the projectile could interact with several nucleons simultaneously (like the one in coherent nucleon tube). The coherent tube interaction with hadrons could give some additional transparency too. The influence of these effects on the nuclear transparency could be studied using two particles correlation technique in our experiment. Leading particles are the high energy projectiles. These high energy projectiles lose a part of their energy during interactions with the target medium. Only a fraction of the energy is transferred to the target because the projectile spent less time in its vicinity and could save other essential part of their energy. Such particles will have maximum energy in an event, small angle (less than $25^0$), which passes very fast by the medium. The particles cannot interact more and that is why medium seems transparent to them. Transparency of these leading protons does not give any information about the properties of the medium due to the reason explained above. The leading effect is suppressed for Cascade Model. It can be explained with some mixture of the fast $\pi^+$-mesons among the leading protons in the experiment. The fast $\pi^+$-mesons could appear as a result of proton exchange reactions when leading proton transform to neutron and fast positive pions appears through the reactions $p \rightarrow n + \pi^+$. Next we plan to study the same average characteristics of pions under the same conditions with the hope that we will get some information about the particular properties of the nuclear matter.